\documentclass[conference,compsoc]{IEEEtran}

\usepackage{titlesec}
\titlespacing\section{0pt}{6pt plus 4pt minus 2pt}{3pt plus 2pt minus 2pt}
\titlespacing\subsection{0pt}{4pt plus 4pt minus 2pt}{3pt plus 2pt minus 2pt}
\titlespacing\subsubsection{0pt}{4pt plus 4pt minus 2pt}{3pt plus 2pt minus 2pt}

\IEEEoverridecommandlockouts
\usepackage{amsmath,amssymb,amsfonts}
\usepackage{marvosym}
\usepackage{hyperref}
\usepackage[linesnumbered,ruled,vlined]{algorithm2e}%[ruled,vlined]{  
\usepackage{algpseudocode} 
\usepackage{xcolor}
\usepackage{graphicx}
\usepackage{subcaption}
\ifCLASSOPTIONcompsoc
  \usepackage[nocompress]{cite}
\else
  \usepackage{cite}
\fi
\ifCLASSINFOpdf
\else
\fi
\begin{document}
\title{MSARS: A Meta-Learning and Reinforcement Learning Framework for SLO Resource Allocation and Adaptive Scaling for Microservices
}
% \thanks{This work is supported by the National Natural Science Foundation of China (No. 62102408),  Guangdong Basic and Applied Basic Research
% Foundation (No. 2024A1515010251), Shenzhen Industrial Application Projects of undertaking the National key R \& D Program of China (No. CJGJZD20210408091600002), and Chinese Academy of Sciences President's International Fellowship Initiative (Grant. 2023VTB0005).}


\author{\IEEEauthorblockN{Kan Hu$^{1, 2}$, Linfeng Wen$^{1, 2}$, Minxian Xu$^{1}$\textsuperscript{\Letter}, Kejiang  Ye$^{1}$}
\IEEEauthorblockA{1. Shenzhen Institute of Advanced Technology, 
Chinese Academy of Sciences, China\\
2. University of Chinese Academy of Sciences, China\\
\{k.hu, lf.wen, mx.xu, kj.ye\}@siat.ac.cn}}


% compsoc conferences compared to traditional conferences), use this
%Telephone: (800) 555--1212, Fax: (888) 555--1212}


\maketitle

\begin{abstract}
Service Level Objectives (SLOs) aim to set threshold for service time in cloud services to ensure acceptable quality of service (QoS) and user satisfaction.
 % Cloud service users only focus on their QoS, while cloud service providers conservatively allocate excessive system resources to ensure services to meet SLOs. 
Currently, many studies consider SLOs as a system resource to be allocated, ensuring QoS meets the SLOs. 
 % However, quickly allocating SLOs to each microservice within a complete service to achieve an optimal resource allocation scheme remains a significant challenge. 
Existing microservice auto-scaling frameworks that rely on SLO resources often utilize complex and computationally intensive models, requiring significant time and resources to determine appropriate resource allocation. 
This paper aims to rapidly allocate SLO resources and minimize resource costs while ensuring application QoS meets the SLO requirements in a dynamically changing microservice environment. 
We propose MSARS, a framework that leverages \underline{m}eta-learning to quickly derive \underline{S}LO resource \underline{a}llocation strategies and employs \underline{r}einforcement learning for adaptive \underline{s}caling of microservice resources. 
It features three innovative components:
First, MSARS uses graph convolutional networks to predict the most suitable SLO resource allocation scheme for the current environment. 
Second, MSARS utilizes meta-learning to enable the graph neural network to quickly adapt to environmental changes  ensuring adaptability in highly dynamic microservice environments. 
Third, MSARS generates auto-scaling policies for each microservice based on an improved Twin Delayed Deep Deterministic Policy Gradient (TD3) model. The adaptive auto-scaling policy integrates the SLO resource allocation strategy into the scheduling algorithm to satisfy SLOs. Finally, we compare MSARS with state-of-the-art resource auto-scaling algorithms that utilize neural networks and reinforcement learning, MSARS takes 40\% less time to adapt to new environments, 38\% reduction of SLO violations, and 8\% less resources cost.
\end{abstract}

\begin{IEEEkeywords}
Microservices, SLO allocation, Meta learning, Reinforcement learning, Resource auto-scaling
\end{IEEEkeywords}


% \fi
\IEEEpeerreviewmaketitle

\section{Introduction} \label{introduction}

Microservices architecture is currently the most popular paradigm in cloud computing \cite{I1}, widely adopted by leading cloud companies such as Google \cite{I7}, Amazon \cite{I6}, and Alibaba \cite{I8}. This architecture decouples large monolithic applications into multiple independent services, each responsible for a specific business function. 
These microservices can be developed, tested, deployed, and scaled independently, allowing for greater flexibility and efficiency in managing cloud applications. However, the dynamic and distributed nature of microservices presents significant challenges for efficient resource scheduling \cite{wang}. Ensuring that these applications perform optimally under varying conditions is critical for maintaining service quality \cite{b5}.

SLOs are goals that specify the desired performance for a service. By using SLOs as a reference for resource allocation, cloud systems can align their resource management strategies with the expectations of users, thereby prioritizing user satisfaction \cite{Parslo}. SLO-based resource allocation ensures that system resources are distributed in a way that meets predefined performance criteria, leading to improved resource utilization, better handling of dynamic workloads, and an overall enhanced user experience.

In microservices-based applications, each microservice is independently deployed and automatically scaled. The latency SLOs are defined for end-to-end services, as end-to-end response time determines user experience. Consequently, in a microservices environment, it is unclear which microservices need to be scaled when end-to-end latency SLO are violated or underutilized, and how to adjust the scale of each microservice at minimal cost to meet the end-to-end latency SLO \cite{Erms}.

Most large systems impose partial latency SLOs on individual microservices along the complete microservice chain, ensuring that if all partial latency SLOs are met, the end-to-end SLO will also be satisfied \cite{Parslo}. Current research has proposed some static end-to-end SLO resource allocation methods, which use Load-Latency Profiles (LLP) graphs depicting the relationship between the number of requests to each microservice and their response time, and related resource gradient descent calculations. These methods can compute approximately optimal SLO resource allocations based on the current environment, allowing each microservice to be assigned the most appropriate partial SLO as a standard for auto-scaling.

\begin{figure*}[htbp]
\centering
\begin{subfigure}[h]{0.245\linewidth}
  \centering
  \includegraphics[width=\linewidth]{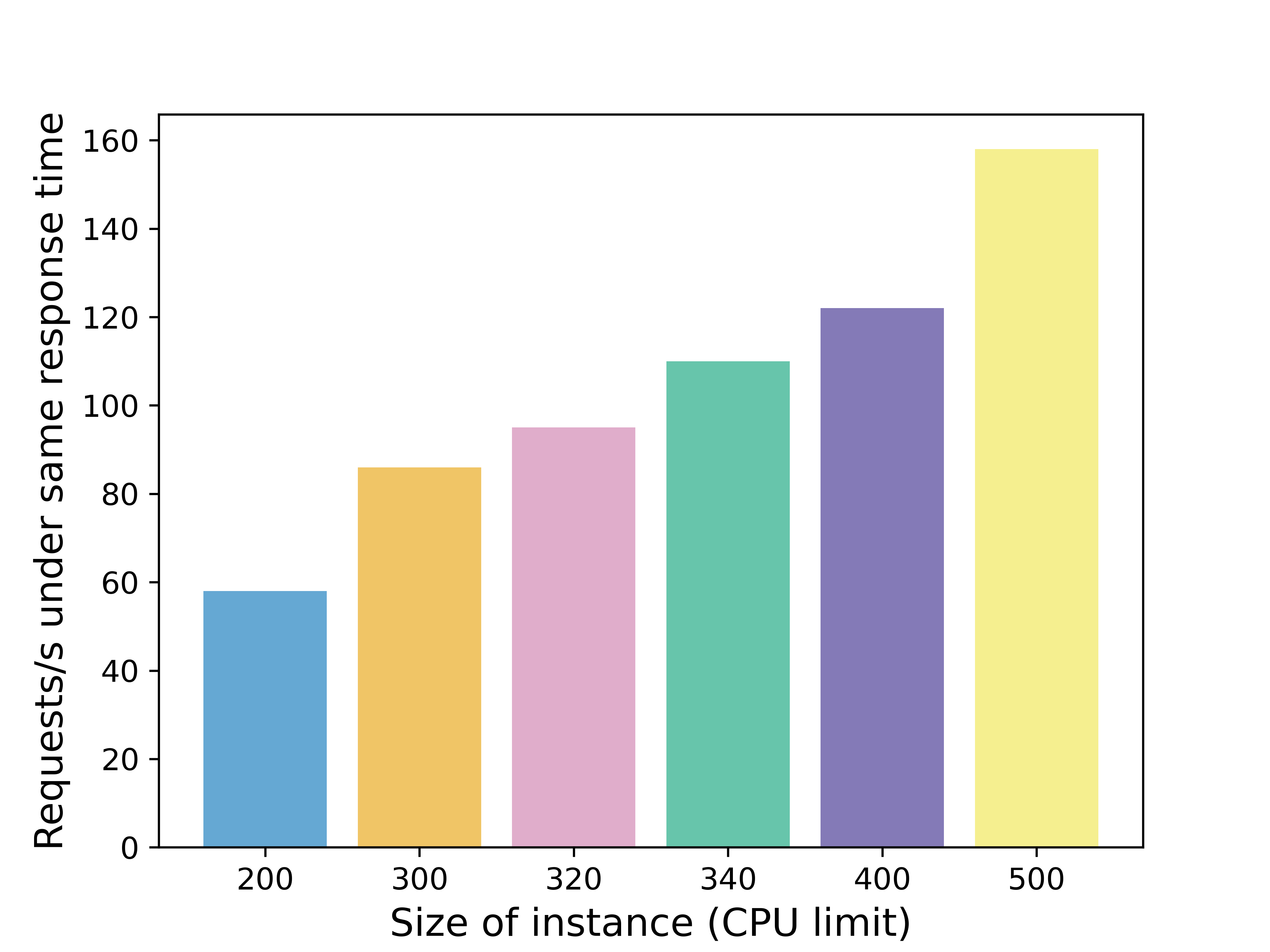}
  \caption{}
  \label{fig:rps}
\end{subfigure}
\begin{subfigure}[h]{0.245\linewidth}
  \centering
  \includegraphics[width=\linewidth]{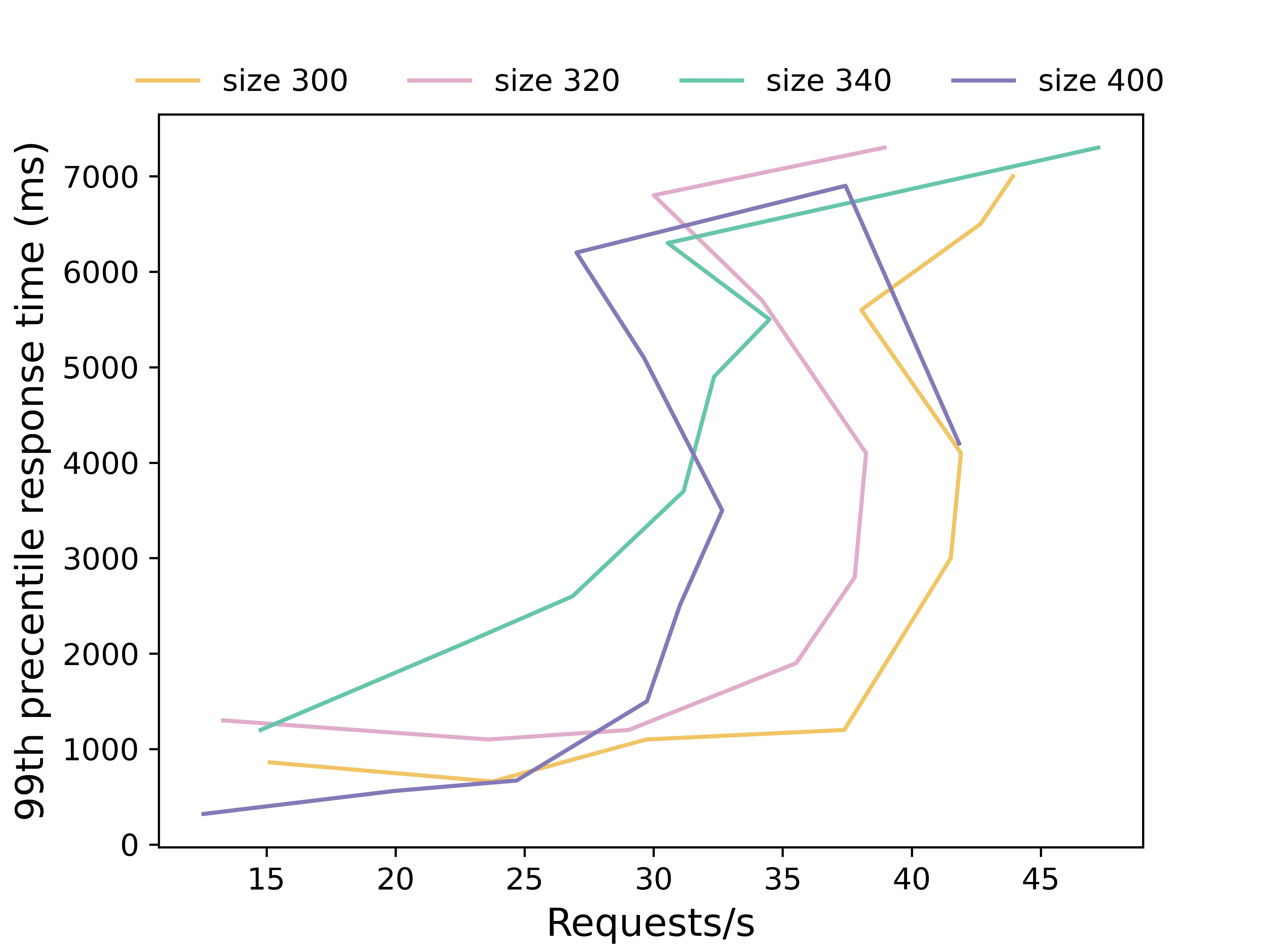}
  \caption{}
  \label{fig:instance_size}
\end{subfigure}
% \caption{}
\begin{subfigure}[h]{0.245\linewidth}
  \centering
  \includegraphics[width=\linewidth]{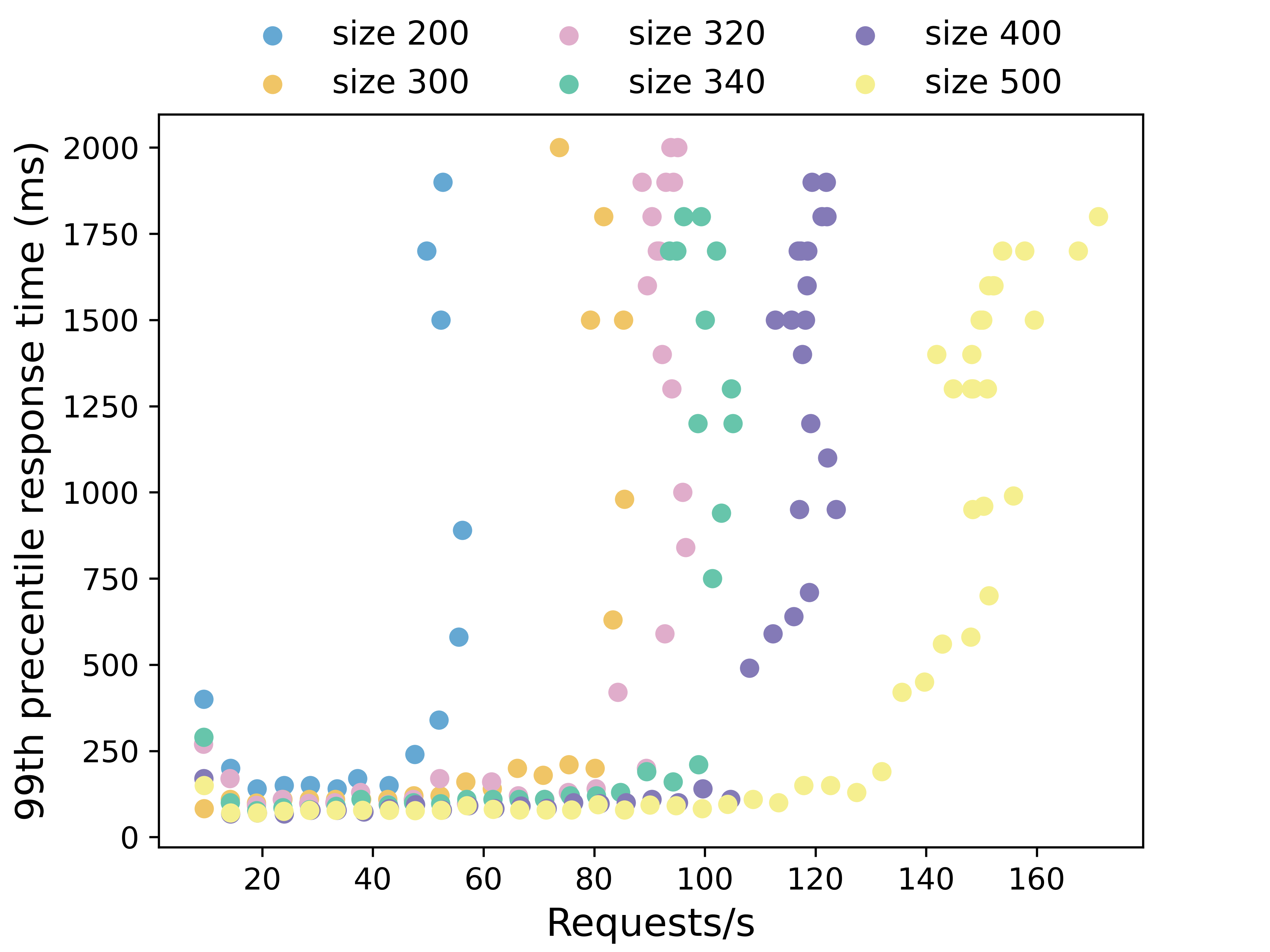}
  \caption{}
  \label{fig:user}
\end{subfigure}
\begin{subfigure}[h]{0.245\linewidth}
  \centering
  \includegraphics[width=\linewidth]{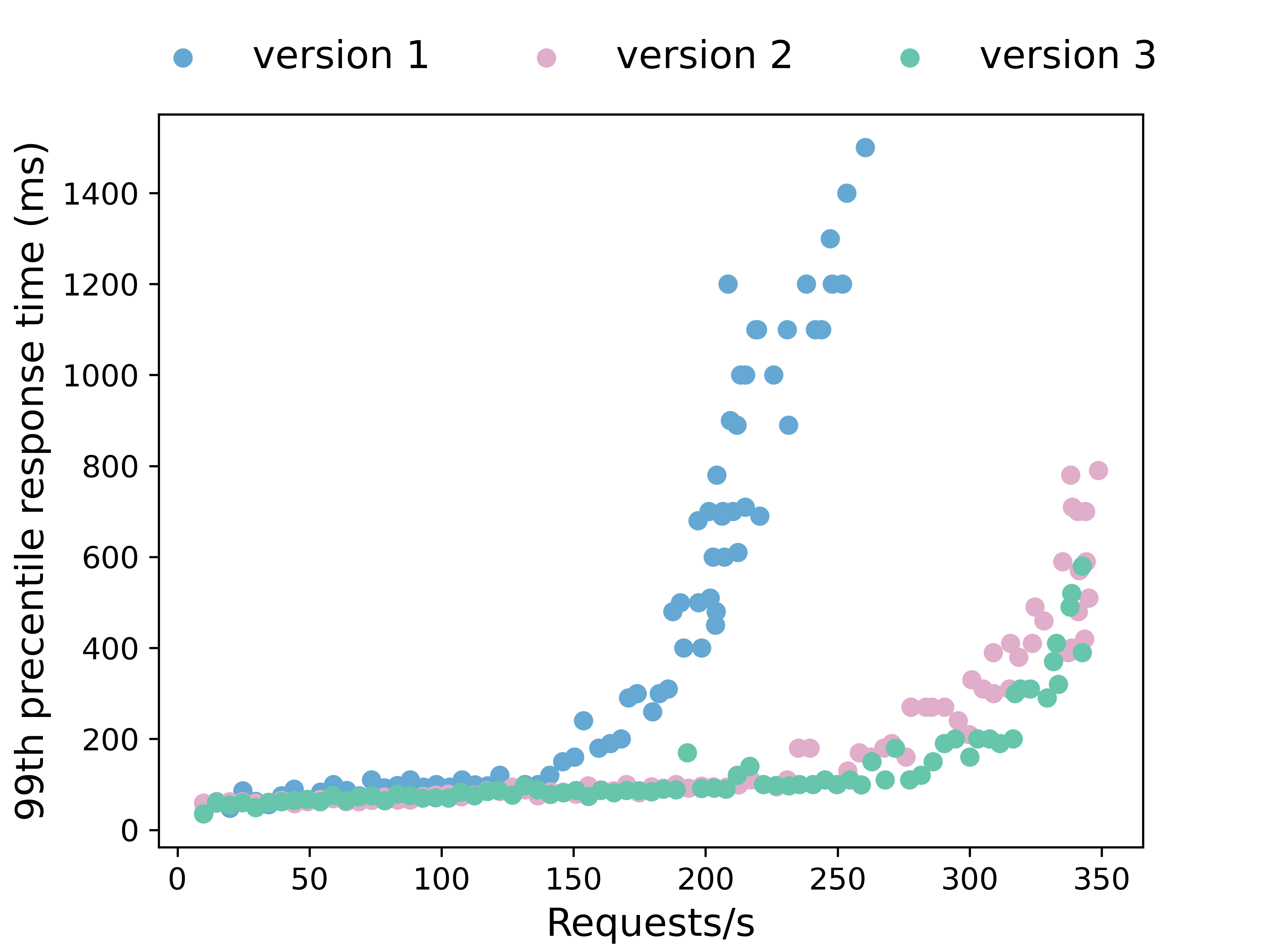}
  \caption{}
  \label{fig:version}
\end{subfigure}
\caption{(a) Capacity of instance to handle concurrent requests with different sizes; (b) LLP for different instance sizes and under different loads; (c) LLP for different size instances; (d) LLP for different versions.}
\label{fig:cap}
\end{figure*}
\vspace{-1pt}

However, these methods heavily rely on a stable environment, consistent applications, and fixed-size container instances \cite{hyscale}. When microservice applications undergo change (such as upgrade) or the size of microservice container instance change, the LLP graphs for these microservices also change. This necessitates updating the LLP for each microservice and obtaining a new partial SLO allocation scheme. Moreover, when tail latency is used as the SLO metric, the non-additive nature of tail latency requires significant computational time to iteratively calculate appropriate partial SLO resource allocations. To address these challenges, and adapt to the high uncertainty of SLO resource allocation in dynamic environments, we apply meta-learning to rapidly adjust to these changes. 

In this paper, we propose SLO-aware resource allocation model that can rapidly adapt to changes in microservices applications and system environments. We also design a reinforcement learning (RL)-based auto-scaling framework to manage resource scaling and ensure service quality within the defined SLOs. The main \textbf{contributions} are as follows:

\begin{itemize}
    \item We analyze the impact of various factors on the load characteristics of microservices applications and their effect on SLO resource allocation. 
    \item [$ \bullet $] We employ a meta-learning model to train SLO resource allocation results for different LLP graphs, various microservice chain structures, and different system resource conditions. This approach allows for the rapid generation of highly uncertain SLO resource allocation schemes, providing optimal scaling thresholds for microservices.
    \item [$ \bullet $] We design a RL-based auto-scaling framework that continuously adjusts resource scaling thresholds. This framework leverages the dynamically updated local SLOs from the meta-learning model to ensure optimal performance and resource utilization.
\end{itemize}

\section{Background and Motivation} \label{motivation}
% Allocating system resources based on SLOs relies heavily on the Load-Latency Profile (LLP) graph of microservices applications. The LLP graph represents the relationship between service request arrival rates and request response times (typically, the 99th percentile response time, also known as tail latency, is used to meet user service quality requirements), as illustrated in Figure 1. By analyzing the LLP graph and the chains structure of microservices, an approximately optimal SLO resource allocation scheme can be determined. However, this method depends on a statically stable system environment; any changes in the LLP graph of a microservice can impact the entire SLO resource allocation outcome. Furthermore, SLO resource allocation requires iterative calculations to derive the approximately optimal partial SLO values.

When using SLO metrics as a system indicator for resource allocation, if the average response time of requests is used to measure SLO resources, the additivity of average response time allows for easy and accurate allocation of SLO resources to each microservice component in the calling graph. However, average response time cannot represent the overall quality of user service. Therefore, the 99th percentile response time, also called tail latency, is usually used as the SLO resource measure. When using this metric, as tail latency is not an additive metric, it is difficult to accurately allocate the SLO resources to each microservice component. Parslo \cite{Parslo} attempts to use an iterative testing model to achieve a near-optimal SLO resource allocation plan. It starts with the current tail latency SLO value as the initial allocation of resources, and uses the gradient descent method to derive SLO resource allocation plan. This is followed by the utilization of its designed offline training and online analysis model to continuously adjust the initial resource allocation, ensuring that the actual link test tail latency equals the tail latency SLO value, thereby obtaining a near-optimal SLO resource allocation plan. However, Parslo requires a significant amount of computation and testing time initially to obtain this allocation plan. 
% Additionally, Parslo relies on the LLP of microservice applications: a graph representing the relationship between service request arrival rate and request response time (usually selecting the top 99\% response time) as shown in Figure~\ref{fig:overall}. Parslo needs to analyze the LLP graph of microservice applications and the structure of the chains, and through continuous offline and online testing, it can obtain a near-optimal SLO resource allocation plan. 
Furthermore, this method relies on a static and stable system environment. Once the LLP graph of a microservice changes, it will affect the SLO resource allocation results for the entire chains. Many factors affect the LLP graph, including the size of microservice application instances, functional attributes of the applications, and the upgrades and assembly of microservice applications. Therefore, we specifically studied the impact of different conditions on the LLP graph.

\subsection{Impact of the Instance Size of Microservices}

The capacity of microservice instances to handle concurrent requests is closely related to their size (allocated resources) \cite{instance}. Generally, larger instances can handle more concurrent requests, as shown in Figure~\ref{fig:rps}, containers of different sizes can handle different maximum numbers of concurrent requests under the same SLO. But there are alternative microserices whose ability to handle concurrent requests does not regularly increase with the size of instances, but instead exists in an irregular manner, as shown in Figure \ref{fig:user}. Therefore, the LLP graphs of microservice container instances of different sizes also have significant differences,
% This is not only reflected in the different capacities to handle concurrent requests but also in the shape of the LLP graphs when plotted against arrival rate, 
as shown in Figures~\ref{fig:instance_size}. Different LLP graphs will lead to markedly different SLO resource allocation results. If we aim to finely control the resources of microservice instances through vertical scaling, this variability can have disastrous effects on our SLO allocation outcomes.

In previous research on SLO resource allocation, to maintain a static LLP graph for microservices and ensure a fixed SLO allocation, each microservice container instance was set to a fixed size, completely abandoning vertical scaling. This method scales microservice resources from a macro perspective, leading to low resource utilization efficiency when handling low traffic loads.

\subsection{Impact of the microservice updates}

Due to the inherent characteristics of microservices, such as containerization, the implementation of version upgrades has become a regular and frequent practice, aimed at meeting ever-evolving business requirements. Microservice version upgrades typically have two notable characteristics. First, these upgrades are frequent and irregular, filled with randomness, often accompanying the introduction of new business requirements, features, or bug fixes. Second, the impact of these upgrades on microservice performance patterns is significantly uncertain. Some upgrades can dramatically increase or decrease the microservice's concurrent request handling capacity, sometimes by more than 100\%, leading to substantial changes in the LLP graph. This could make the LLP steeper, allowing the microservice to handle more requests under the same SLO, or it could make the LLP flatter, we have a test as shown in Figure \ref{fig:version}.

There are many scenarios for microservice version upgrades, which can be broadly categorized into four types: optimizing application performance, fixing bugs, adding new features, and adding new functionalities. Interestingly, the impact of these four types of upgrades on the microservice LLP is inconsistent and highly dependent on the quality of the underlying running code logic during the upgrade. However, for microservice managers, the operational logic within the container is entirely transparent, making it impossible to estimate the impact of the next version upgrade on the LLP or its effect on SLO resource allocation.

However, upgrades in microservice applications are frequent and irregular. Previous methods required a static environment and continuous feedback testing to obtain the optimal SLO resource allocation scheme, which is impractical for realistic production. Therefore, a method that can quickly adapt to these unpredictable version updates is required.

\subsection{Meta-learning and Reinforcement Learning}

Meta-learning \cite{meta24}, known as “learning to learn,” is a subfield of machine learning focused on developing algorithms that can quickly adapt to new tasks by leveraging prior experience from related tasks. In the realm of microservices resource scheduling, meta-learning can facilitate rapid adaptation to new application characteristics or workload patterns. This capability is crucial for maintaining optimal performance in environments where the operational context changes frequently. By leveraging meta-learning \cite{maml}, systems can generalize from past experiences to make informed decisions about resource allocation, even when faced with novel scenarios.

To ensure that microservices applications can automatically scale while maintaining service quality during operation, we integrated and improved our previous work on a RL-based autoscaler \cite{DRPC}. This enhanced autoscaler can accept partial SLOs obtained through meta-learning as thresholds for horizontal scaling. It also returns the results of vertical scaling to the meta-learning algorithm, which then recalculates suitable partial SLO values for the current environment and container instance sizes.

% By incorporating meta-learning, the system can rapidly adapt to changes in the application or environment, ensuring that resource allocation remains optimal. The RL-based autoscaler dynamically adjusts the scaling actions based on the updated local SLOs, thereby maintaining the desired performance levels and optimizing resource usage. This integrated approach leverages the strengths of both RL and meta-learning to handle the complexities and uncertainties of microservices resource management.

\begin{algorithm}[t]
\setlength{\abovecaptionskip}{2pt} % 调整伪代码标题上方的距离
\setlength{\belowcaptionskip}{2pt} % 调整伪代码标题下方的距离
\DontPrintSemicolon
\SetAlgoLined
\KwResult{Ensure QoS within SLO constraints and minimise resource consumption}

Initialize instance size  $S_i$, chains structure $L$ and node load-latency-profile parameters $p_1$, $p_2$\;
Initialize task of Meta Learner $T_i$\;
Initialize GCN model parameter set $\theta$ by Meta Learner\;
\While {Service monitor executing}{
    Meta $\gets$ Edge:$A$, Node:$X$, Global:$g$\;
    Workload Forecaster $\gets$ Current load: $CL$\;
    RL agent $\gets$ Future load: $FL$, State: $s$, SLO: $SLO_{\text{partial}}$\;
    RL agent generate Scaling decision\;
    Scaling execution\;
    \If{Execute vertical scaling}{
        Meta $\gets$ instance size $c$\;
        Meta $\gets$ Edge:$A$, Node:$X$, Global:$g$\;
        RL agent $\gets$ SLO: $SLO_{\text{partial}}$ \;
    }    
    }
\caption{MSARS: Workflow.}
\label{alg:workflow}
\end{algorithm}

\begin{figure}[htbp]
\centering
\includegraphics[width=1.0\linewidth]{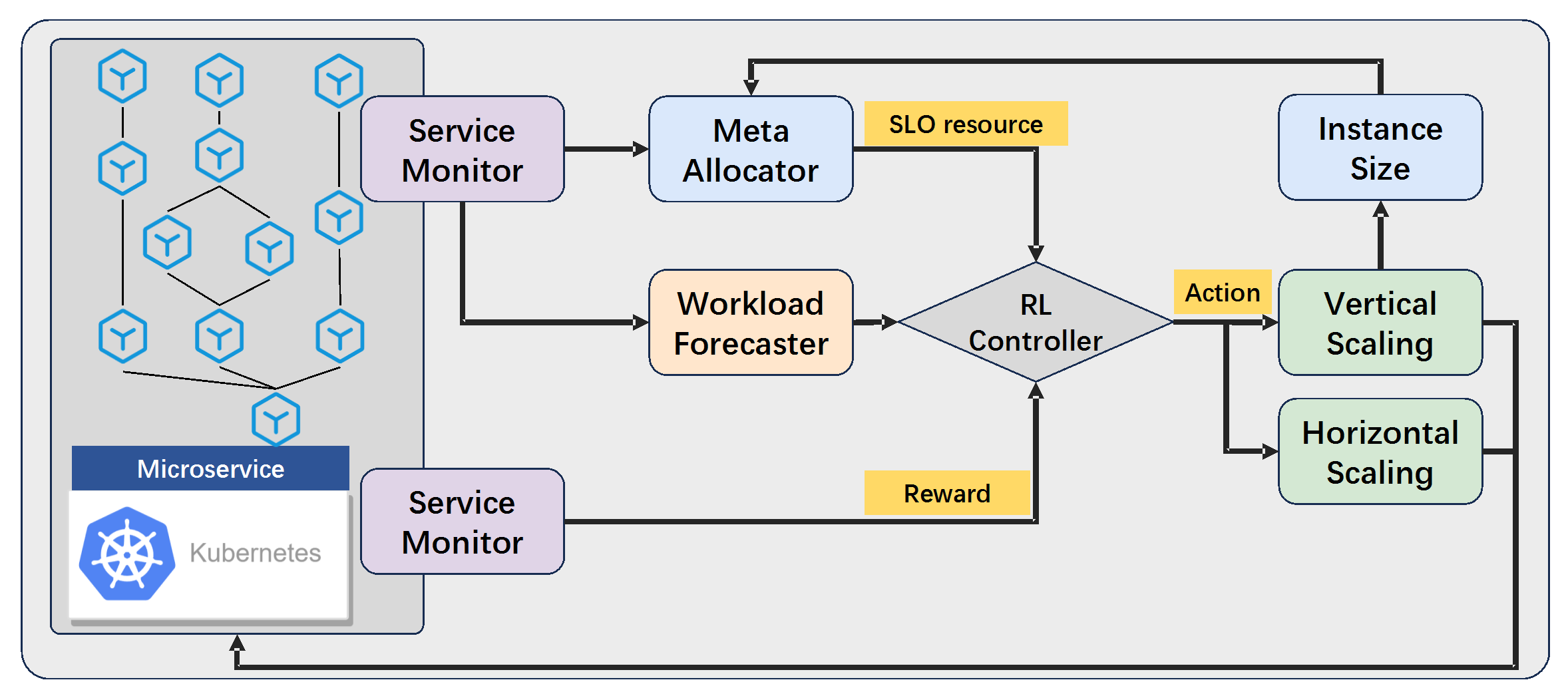}
\caption{Workflow of MSARS.}
\label{fig:overall}
\end{figure}

\section{MSARS Workflow Design}

% \subsection{Overview}

To address the issues of rapid SLO resource allocation, ensuring user service quality, and automatically managing system resources in a dynamic microservice-based environment, we design an auto-scaling management framework for microservices, MSARS, based on meta-learning and supporting RL agents. MSARS quickly analyzes and derives new SLO resource allocation plans based on changes in the LLP graph of microservice applications. When new applications or LLP graphs appear in the system environment, MSARS can quickly adapt and learn from these changes, ensuring that it always provides the most resource-efficient SLO resource allocation plan. Figure~\ref{fig:overall} provides an overview of MSARS and the workflow. The overall workflow of MSARS is shown in Algorithm~\ref{alg:workflow}.

\subsection{Service Monitor}
The \textit{Service Monitor} is the top-level perception component in the MSARS architecture, used to monitor the entire cloud service cluster. It collects real-time metrics of all services, including current workload status, microservice chain structure, each service's LLP graph, request response times, instance sizes and numbers, and CPU utilization. The \textit{Service Monitor} periodically checks the LLP graph of microservices. When changes in the LLP graph occur, it sends the new changes combined with the chain structure to \textit{Meta Allocator} for reallocation of SLO resources. It monitors the current workload status and sends it to the workload forecaster to predict the next stage of workload. It also monitors request response times, instance sizes and numbers, and CPU utilization, sending these as State in Algorithm~\ref{alg:workflow} to the \textit{RL Controller}.

\subsection{Meta Allocator}
To quickly obtain an accurate near-optimal SLO resource allocation plan, we design a Graph Convolutional Network (GCN) structure to analyze the relationships between chain structure, LLP graph, and SLO resource allocation. Due to the irregular changes in LLP graphs and the rich diversity of chain structures of microservices, accurately achieving SLO allocation with the GCN model in new scenarios is challenging in practice. Therefore, we utilize a meta-learner to address this challenge. The meta-learner can quickly adapt to environmental changes with small sample training. When significant changes occur in the system environment, the meta-learner can detect this shift and generate new graph neural network parameters to adapt to these changes. \textit{Meta Allocator} is the core component of MSARS, ensuring that the entire architecture quickly adapts to  unfamiliar environments and generates accurate SLO resource allocation.

\subsection{Workload Forecaster}
This module receives the current workload status from the \textit{Service Monitor} and uses a gated recurrent unit model to predict the workload for the next time segment. Its output is sent to the \textit{RL controller} as the State in Algorithm~\ref{alg:workflow}, guiding the \textit{RL Controller}'s next Action in Algorithm~\ref{alg:workflow}. This module leverages our previous work, the esDNN model \cite{esdnn}, a deep learning network based on supervised learning. This model is lightweight and accurately predictive, providing reliable workload predictions for our framework.

\subsection{RL Controller}
In this module, a RL model is adopted to generate auto-scaling policies. This model is based on the TD3 model \cite{td3}, makes decisions by receiving monitored request response times, instance sizes and numbers, CPU utilization from the \textit{Service Monitor}, and the next time slot's workload prediction from the \textit{Workload Forecaster}. Meanwhile, \textit{Meta Allocator} provides the generated SLO resource allocation plan to the controller, serving not only as an input State but also as a critical factor in its auto-scaling decisions.

\section{Framework of MSARS}
\begin{figure*}[htbp]
\centering
\includegraphics[width=1.0\linewidth]{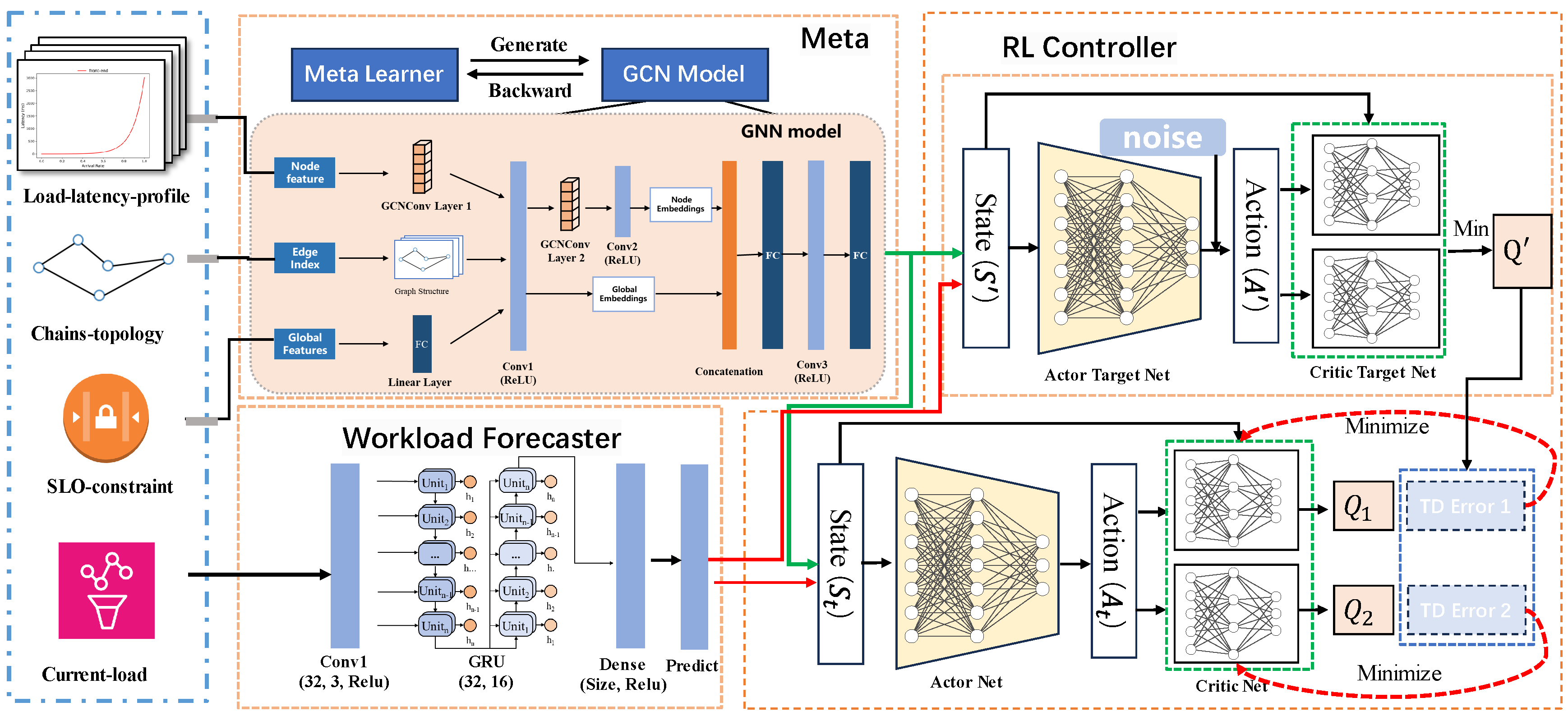}
\caption{MSARS framework that can quickly handle changes in microservices based on meta-learning and RL.}
\label{fig:Core}
\end{figure*}

In this section, we will detail the functioning of our core modules in MSARS framework, as shown in Figure~\ref{fig:Core}.

\subsection{Meta Allocator}
\textit{Meta Allocator} is responsible for generating SLO resource allocation strategies. To accurately capture the impact of microservice chain structure and LLP graph on SLO resource allocation, we develop a GCN model. This model effectively captures the relationships between nodes and edges by defining convolution operations on graph-structured data. Additionally, to adapt to the complex and dynamic chain structures of microservice applications, the varying LLP graph characteristics, and the different SLO constraints of different applications, we designed a \textit{Meta Learner}. \textit{Meta Learner} enables the GCN model to quickly adapt to new environments, ensuring the effectiveness of the SLO resource allocation strategy.

\subsubsection{GCN Model} The goal of the GCN model is to generate accurate and effective SLO resource allocation plans based on end-to-end SLO resource constraints combined with the microservice chain structure and the nodes' LLP graphs. We decompose the chain structure of microservices into a directed acyclic graph with a single end-to-end SLO and input it into the model as an Edge Index in the form of an adjacency matrix $ A \in \mathbb{R}^{N \times N}$. The LLP of each microservice component node is input into the model as a Node feature in the form of a node feature matrix $ X \in \mathbb{R}^{N \times D}$, where $N$ denotes the number of nodes and $D$ denotes the dimension of node features. The end-to-end SLO constraint is input as a Global feature in the form of a feature vector $g$ with dimensions $ 1 \times F$, where $F$ indicates the dimension of global feature. 

To ensure that nodes retain their feature information during subsequent convolution operations, we first add self-loops in the adjacency matrix $A$, resulting in $\tilde{A} = A + I$, where $I$ is the identity matrix. We then compute the degree matrix $\tilde{D}_{ii} = \sum_j \tilde{A}_{ij}$. To avoid changes in the scale of feature vectors, we use symmetric normalization to process the adjacency matrix, yielding $\hat{A} = \tilde{D}^{-\frac{1}{2}} \tilde{A} \tilde{D}^{-\frac{1}{2}}$.

In the GCN layer, we perform convolution calculations on the node features and the normalized adjacency matrix. The first convolution layer transforms the input node feature matrix $X$  into a hidden representation $H^{(1)}$,
\begin{equation}
    H^{(1)} = \sigma \left( \hat{A} X W^{(0)} \right),
\end{equation}
\begin{equation}
    H^{(2)} = \sigma \left( \hat{A} H^{(1)} W^{(1)} \right),
\end{equation}
where $W$  is the weight matrix of the first layer and $\sigma$ is the activation function, which is ReLU in our model. The second convolution layer transforms the first hidden representation $H^{(1)}$ into a deeper hidden representation $H^{(2)}$.

The hidden representations can be abstracted as:
\begin{equation}
   H^{(l+1)}_i = \sigma \left( \sum_{j \in \mathcal{N}(i) \cup \{i\}} \frac{1}{\sqrt{\tilde{D}_{ii} \tilde{D}_{jj}}} H^{(l)}_j W^{(l)} \right).
\end{equation}

We process the global feature vector $g$ through a fully connect layer to obtain the global feature embedding $g_{emb}$: 
\begin{equation}
    g_{emb} = \sigma \left( g W_g \right),
\end{equation}
 where $W_{g}$ is the weight matrix of the fully connected layer. 
 % In order to better combine the global characteristics end-to-end SLO constraints with the node characteristics of the microservice nodes to analyse the relationship between end-to-end SLOs and local SLO resource allocation, 
 To better capture node information and global graph structure to analyse the relation between SLO and LLP parameters, we combine the node features and global features. We merge the output $H^{(2)}$ of the second convolution layer with the global feature embedding $g_{emb}$. Since $H^{(2)}$ is a node feature matrix with dimensions $N \times D'$, where $D'$ is the dimension of hidden layer, 
 % and $g_{emb}$ is a global feature vector with dimensions $1 \times D'$, 
 we need to scale $g_{emb}$ to match the shape of $H^{(2)}$:
 \begin{equation}
     g_{emb\_expanded} = g_{emb}.repeat(N, 1),
 \end{equation}
and then concatenate the node features and global features along the feature dimension: 
\begin{equation}
H_{combined} = [H^{(2)} \parallel g_{emb_expanded}].
\end{equation}
We input $H_{combined}$ into a fully connected layer to further integrate the feature information within the model, producing the final output: 
\begin{equation}
O_{output} = \sigma(H_{combined} W_{out}).
\end{equation}
We use pre-tested end-to-end SLO resource allocation schemes for tail latency as the dataset for the entire model, ensuring that the output of the GCN model closely approximates the ideal local SLO resource allocation scheme.

GCN model usually has the problem of consuming a lot of computational resources and time due to the increase in the number of nodes, to address this, we decouple all the microservice chains into simple end-to-end microservice chains, and the GCN model only needs to compute the end-to-end microservice chains that have been changed each time, which has the benefit of a smaller number of nodes and a simpler chain, and ensures the accuracy of the GCN model prediction and the real-time.

\subsubsection{Meta Learner} In the context of microservice architecture applications, different microservice applications exhibit significant variations in microservice chains structures, microservice component node LLPs, and end-to-end SLO constraints. Whenever a new microservice application is encountered, or the LLP graph of a microservice application undergoes significant changes, the accuracy of the model's predictions can be greatly reduced. To enable our GCN model to quickly adapt to new tasks, we design a \textit{Meta Learner}. The aim is to train the model to quickly adapt to new tasks by training on multiple tasks, enabling the model to perform well on new tasks after only a few gradient updates. Our goal is to find an initial parameter set $\theta$ through the \textit{Meta Learner} component, where $\theta$ represents the relevant training parameters in the GCN model, including the weights $W^l$ of the GCN convolution layers, the weights $W$ and biases $b$ of the fully connected layers. This ensures that when given a new task, the model can quickly obtain the model parameters $\theta'$ adapted to that task through a few gradient updates \cite{maml}.

Given the differences between microservice applications, we use the microservice chain structure as the task division standard. An end-to-end chain of a microservice application is considered as a task division for training. By controlling the microservice component nodes, we alter the LLP of the microservice component nodes and different end-to-end constraints to form the dataset $D$ of a single task $T_i$. This dataset is divided into the training set $D^{train}_{i}$ and the test set $D^{test}_{i}$, with 70\% of dataset used as the training set and 30\% as the test set in this experiment. Different end-to-end chains of different microservice applications constitute different task datasets.

The entire meta-learning process can be divided into the inner loop and the outer loop.  

\textbf{Inner Loop} involves task-level training. For each task $T_i$, we use the current model parameter set $\theta$ to perform gradient descent, updating the task-specific parameters $\theta_{i}'$.  For each task $T_i$, we compute the gradient of the loss: $\nabla_\theta \mathcal{L}{\mathcal{T}i^{\text{train}}} (\theta)$, and use this gradient to update the model:

\begin{equation}
\theta_i' = \theta - \alpha \nabla\theta \mathcal{L}{\mathcal{T}_i^{\text{train}}} (\theta),
\end{equation}

where $\alpha$ denotes the learning rate.

\textbf{Outer loop} involves meta-level training, calculating the loss on the test sets of all tasks, and optimizing the initial parameters $\theta$. First, we calculate the loss on the test set for each task with the updated parameters $\theta_{i}'$: 

\begin{equation}
\mathcal{L}{\mathcal{T}i^{\text{test}}} (\theta_i')=\mathcal{L}{\mathcal{T}i^{\text{test}}} (\theta - \alpha \nabla\theta\mathcal{L}{\mathcal{T}i^{\text{train}}} (\theta)).
\end{equation}

Then, we compute the meta-gradient for the entire model: $\nabla\theta \sum_{\mathcal{T}i \sim p(\mathcal{T})} \mathcal{L}{\mathcal{T}i^{\text{test}}} (\theta_i')$, and update the initial parameters accordingly:

\begin{equation}
\theta \leftarrow \theta - \beta \nabla\theta \sum_{\mathcal{T}i \sim p(\mathcal{T})} \mathcal{L}{\mathcal{T}_i^{\text{test}}} (\theta_i'),
\end{equation}
where $\beta$ denotes the learning rate.

Through this process, the GCN model obtains a set of model parameters with good generalization capability. When facing new tasks, the GCN model can quickly obtain the model parameters $\theta_{i}'$ adapted to those tasks through a few gradient updates.

\subsection{RL Controller}
MSARS leverages RL to optimize resource management strategies, aiming to achieve long-term rewards in dynamic microservices environments. 
% Due to the dynamic nature of microservices, existing performance modeling methods or heuristic-based methods are troubled by model reconstruction and retraining. 
RL, through a tight feedback loop, continuously explores the action space and generates optimal strategies without relying on predefined assumptions (such as heuristic rules). It can directly learn from actual workloads and environments, understanding the impact of overall resource scaling on microservice QoS. MSARS employs the TD3 algorithm, a model-free, Actor-Critic framework combining policy-based and value-based RL frameworks (as shown in the Figure \ref{fig:Core}). The SLO resource allocation plan generated by the \textit{Meta Allocator} is also included as part of the environment state to guide the RL agent in decision-making.

Additionally, to make MSARS's RL framework more suitable for SLO-driven microservice application systems, we include the real-time updated SLO resource allocation strategy generated by the \textit{Meta Learner} as a part of reward for RL agent. This allows the RL agent to better identify resource scaling thresholds when guiding resource scaling, ensuring QoS while reducing resource consumption.

\subsubsection{Reward Formulation} The goal of MSARS is to maximize system resource utilization while ensuring SLOs. Therefore, our reward model is based on response time and resource utilization, combined with the SLO resource allocation plan generated by the \textit{Meta Learner}, resulting in the following reward function model:

\begin{equation}
    \label{eq:reward}
    R_{qos}(rt) = \begin{cases}
        e^{-\left(\frac{rt - SLO_{\text{partial}}}{SLO_{\text{partial}}}\right)^2}, &  rt > SLO_{\text{partial}} \\
        1 &, rt \leq SLO_{\text{partial}},
    \end{cases}
\end{equation}

where $rt$ denotes the response time (the latency bwtween a request being sent and the return of the result), which measures service quality. $SLO_{partial}$ is the partial SLO resource generated by the \textit{Meta Learner}, representing the maximum delay each microservice component can tolerate. When the response time exceeds $SLO_{partial}$,  it is likely to result in SLO violation. Under normal system operation, a reward of 1 is assigned. However, if performance exceeds $SLO_{partial}$, the reward is penalized and gradually approaches 0, thereby discouraging overloading.

\begin{equation}
    \label{eq:rewardutil_modified}
    R_{util}(v_k) = \begin{cases}
        \frac{\sum_{k=1}^K \sum_{r=1}^R (V_{rk}^{pref} - v_{rk})^3}{K} + 1 &, v_{rk} \leq V_{rk}^{pref} \\
        \frac{\sum_{k=1}^K \sum_{r=1}^R (v_{rk} - V_{rk}^{pref})^3}{K} + 1 &, v_{rk} > V_{rk}^{pref}
    \end{cases}
\end{equation}

Ideal resource utilization is also one of our goals, so we have included system resource utilization in our reward function, as shown in Eq. 2, where $K$ and $R$ represent the $kth$ physical machine and $rth$ resource type residual on the $kth$ physical machine, respectively. $V_{rk}^{pref}$ denotes the preferred utility for a specific resource type $r$ on machine $k$. Proximity to this ideal utility is rewarded, while underprovisioning and wastage are discouraged.

\begin{equation}
	\label{eq:finalreward}
	r(s_t, a_t) = \frac{R_{qos}(rt)}{R_{util}(v_k)}.
\end{equation}

The final reward value (Eq.(\ref{eq:finalreward})) integrates both response time and resource utilization, incentivizing higher $R_{qos}$ values and lower $R_{util}$ values.

The pseudocode of the training algorithm is shown in Algorithm \ref{alg:TD3}, TD3 initiates by initializing actor and critic networks and constructing a replay buffer for storing transition data (lines 1-3). TD3 training proceeds in episodes and each episode consists of $T$ time steps. An action with added exploration noise $\epsilon$ is selected, and the corresponding time-step information is stored in the replay buffer $E$, including $(s,a,r,s',d)$, where $d$ denotes the signal of end. 

\begin{algorithm}[t]
\setlength{\abovecaptionskip}{2pt} % 调整伪代码标题上方的距离
\setlength{\belowcaptionskip}{2pt} % 调整伪代码标题下方的距离
\DontPrintSemicolon
\SetAlgoLined
%\KwResult{TD3 (Twin Delayed Deep Deterministic policy gradient)}
\KwResult{TD3 decisions}

Initialize policy network $\mu$ and value networks $V_{\phi_1}$, $V_{\phi_2}$ with random parameters $\psi, \phi_1, \phi_2$\;
Initialize target networks $\mu'$ and $V'$ with weights $\mu' \gets \mu, V'_{\phi_{1'}} \gets V_{\phi_1}, V'_{\phi_{2'}} \gets V_{\phi_2}$\;
Initialize experience replay buffer $E$\;
\For{episode = 1 to N}{
    Receive initial state $s$\;
    \For{t = 1 to T}{
        Select action $a = \mu(s) + \epsilon$ and execute it\;
        Observe new state $s'$, reward $r$ and done signal $d$ to end episode\;
        Store $(s, a, r, s', d)$ in $E$\;
        \If{it's time to update}{
            Sample a batch $C$ of transitions $(s, a, r, s', d)$ from $E$\;
            $a' \gets \mu'(s') + \text{clip}(\epsilon, -\delta, \delta)$\;
            $y \gets r + \gamma (1 - d) \min_{j=1,2} V'_{\phi_{j'}}(s', a')$\;
            Update $V_{\phi_j}$ by minimizing the loss: $L(\phi_j) = \frac{1}{|C|}\sum_{(s, a, r, s', d) \in C}(V_{\phi_j}(s, a) - y)^2$\;
            \If{$t \mod \text{policy\_update} == 0$}{
                Update $\mu$ by minimizing the loss: $L(\psi) = \frac{1}{|C|}\sum_{s \in C} -V_{\phi_1}(s, \mu(s))$\;
                Soft update the target networks (for both policy and value): $\phi_{j'} \gets \tau \phi_{j'} + (1 - \tau)\phi_j$ \;
            }
        }
        $s \gets s'$\;
    }
}
\caption{MSARS: TD3 Algorithm.}
\label{alg:TD3}
\end{algorithm}

% TD3 training proceeds in episodes and each episode consists of $T$ time steps. Due to the presence of the delayed update strategy, TD3 delays updating the policy network and target network parameters, updating the policy network only every few steps. When the training process reaches an update time, the network will be updated, TD3 randomly samples a batch $B$ of transition records $(s,a,r,s',d)$ from the replay buffer $R$, then uses the target policy network and target Q-value network to compute the target action $a'$: 

TD3 training is organized into episodes, each comprising $T$ time steps. Owing to its delayed update strategy, TD3 updates the policy network and target network parameters only intermittently. At designated update intervals, TD3 samples a batch $C$ of transitions $(s,a,r,s',d)$ from the replay buffer $E$. The target policy network and target Q-value network are then used to calculate the target action $a'$ (lines 11-14):
\begin{equation}
	a' \gets \mu'(s') + \text{clip}(\epsilon, -\delta, \delta).
\end{equation}
Then the target Q-value $y$ is computed by follow function: 
\begin{equation}
	y \gets r + \gamma (1 - d) \min_{j=1,2} V'_{\phi_{j'}}(s', a'),
\end{equation}
where the Q-value is the smaller value from the twin critic networks, effectively addressing the risk of Q-value overestimation. The loss is then calculated based on the Mean Squared Error (MSE), and update the $V_{\phi_j}$ by minimizing the loss:
\begin{equation}
	L(\phi_j) = \frac{1}{|C|}\sum_{(s, a, r, s', d) \in C}(V_{\phi_j}(s, a) - y)^2,
\end{equation}
where $C$ is a random batch of convert records in replay buffer $E$.
Due to the presence of the delayed update strategy. The policy network is only updated when the current time step meets the policy update condition $t\mod \text{policy\_update} == 0 $ (line 15),
\begin{equation}
	L(\psi) = \frac{1}{|C|}\sum_{s \in C} -V_{\phi_1}(s, \mu(s)),
\end{equation}
and soft update the target networks (for both actor and critic) at the same time:
\begin{equation}
	\phi_{j'} \gets \tau \phi_{j'} + (1 - \tau)\phi_j.
\end{equation}

\section{Experimental Evaluations}
In this section, we introduce our experimental settings to evaluate our approach and discuss performance improvement achieved by MSARS in terms of algorithm robustness and SLO violation ratio under dynamic environment.  
\subsection{Experimental Setup}
We evaluate MSARS using Sock-Shop\footnote[1]{https://github.com/microservices-demo/microservices-demo}, an e-commerce website microservice with 10 available versions and 5 main chains. We randomly combined 10 different versions of the microservice application, instances of each microservice with varying sizes, and different end-to-end chains to form the dataset for meta-learning and GCN model training. Each specific version, end-to-end chain, and the SLO resource allocation scheme generated by its microservice instances of different sizes are treated as a separate task for the meta-learner's training. In this way, we obtained a dataset that captures variations in SLO resource allocation schemes due to multiple path structures, different versions, and changes in instance sizes.

Kubernetes is used for underlying container orchestration. We deploy Kubernetes clusters on 10 machines comprising 3 master nodes and 7 worker nodes, the 3 master nodes are equipped with 32 cores and 64GB memory each, among the 7 worker nodes, 4 are configured with 32 cores and 64GB memory, and the remaining worker nodes are with 56 cores and 128GB memory, 104 cores and 256GB memory, 64 cores and 64GB memory. The load generators utilize Alibaba's 2022 dataset\footnote[2]{https://github.com/alibaba/clusterdata/tree/master/cluster-trace-microservices-v2022} to create workloads for sample collection and the creation of traffic surges, operating on a separate machine from the Kubernetes clusters. To remove the randomness, each experiment has been repeated three times.

\begin{figure*}[htbp]
\centering
\begin{subfigure}[h]{0.49\linewidth}
  \centering
  \includegraphics[width=\linewidth]{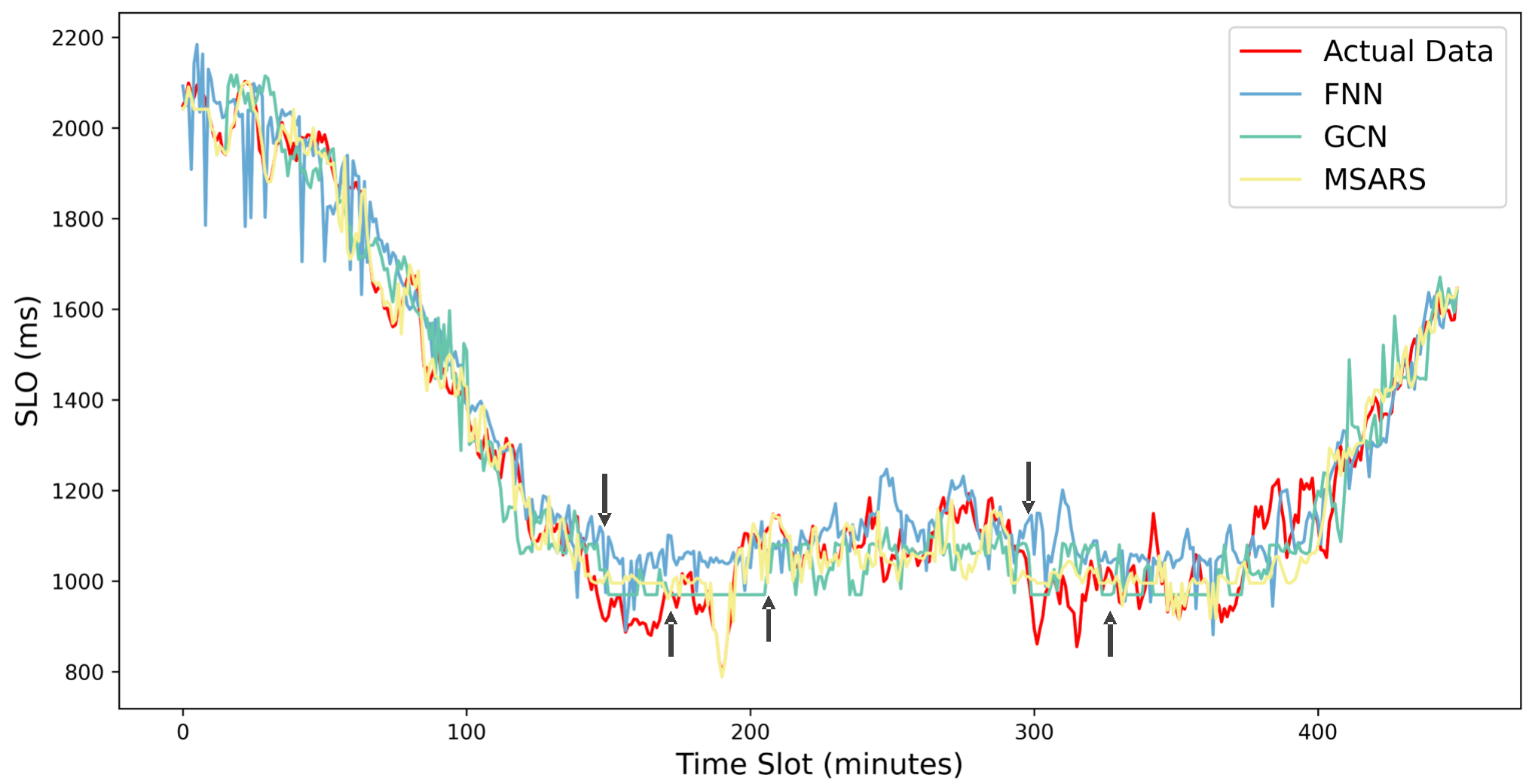}
  \caption{}
  \label{fig:test}
\end{subfigure}
\begin{subfigure}[h]{0.49\linewidth}
  \centering
  \includegraphics[width=\linewidth]{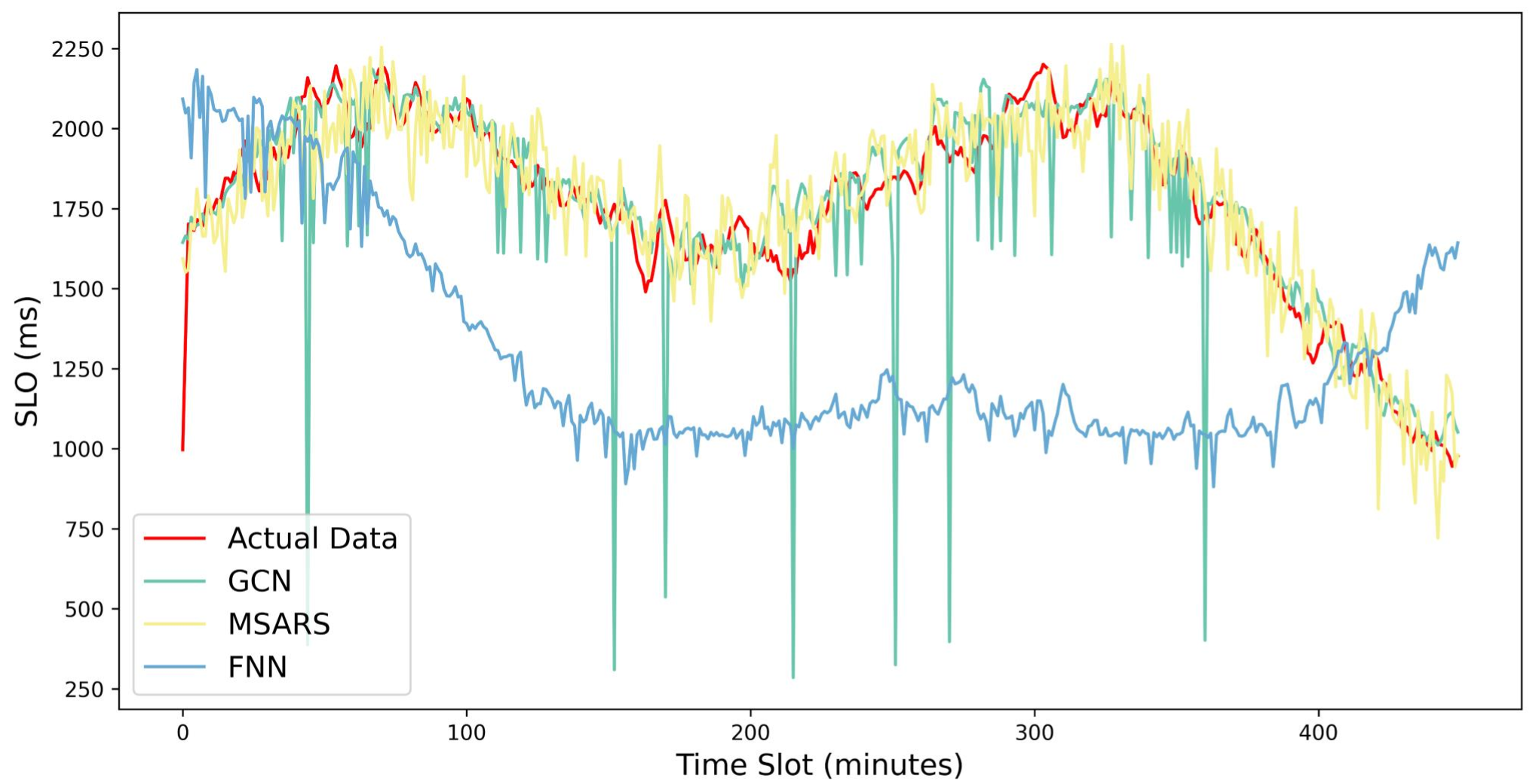}
  \caption{}
  \label{fig:test2}
\end{subfigure}
\caption{(a) Robustness Comparison of MSARS and baselines; (b) Comparison under varied chain structures.}
\label{fig:te}
\end{figure*}

\subsection{Robustness and Accuracy of Meta Allocator}

In order to test the adaptability of our GCN model after being trained by the \textit{Meta-Learner}, we specifically chose test data with chains structures significantly different from the training data and with different LLP graph shapes. We compared it with a convergent GCN model \cite{KRGCN} trained on the training set and a Feedforward Neural Network (FNN) model \cite{fnn}, which is a type of artificial neural network where connections between the nodes do not form a cycle, FNN models can effectively capture non-linear relationships in data and are widely used for tasks such as data classification, regression, and pattern recognition, as shown in the Figure \ref{fig:test}. Compared to the other two baseline algorithms, the GCN model from MSARS, which was trained through \textit{Meta Learner}, is better able to capture changes in SLO resource allocation schemes due to environmental variations, including changes in microservice instance sizes and application version updates. Specifically, around time slot 120 and 300 (each time slot represents 1 minute), we made significant changes to the LLP parameters, causing the predicted results to become distorted. This was particularly evident in the FNN model, which produced ineffective predictions for an extended period until retraining was complete. In contrast, after feedback and rapid parameter updates by the meta-learner, MSARS required less time (more than 40\%) to retrain compared to the other models, quickly generating SLO resource allocation strategies suited to the new environment. Additionally, we observed that the converged GCN model took longer to retrain parameters to maintain the effectiveness of predictions after the model became distorted.

% \begin{figure}[htbp]
% \centering
% \includegraphics[width=1.0\linewidth]{test.png}
% \caption{Robustness Comparison of MSARS and baselines.}
% \label{fig:test}
% \end{figure}

Next, we explored whether MSARS could quickly adapt to new microservice applications, unfamiliar chain structures, and similar LLP parameter combinations with minimal parameter training. As shown in the Figure \ref{fig:test2}, during testing, we found that the FNN model did not recognize changes in microservice chain structures and continued to generate SLO resource allocation schemes based on previously trained parameters. However, MSARS identified these chain structure changes through Edge feature and made corresponding adjustments. The converged GCN model also recognized these changes but took longer to converge.

% \begin{figure}[htbp]
% \centering
% \includegraphics[width=1.0\linewidth]{test2.png}
% \caption{Comparison under varied chain structures.}
% \label{fig:test2}
% \end{figure}

\begin{figure}[htbp]
\centering
\includegraphics[width=0.8\linewidth]{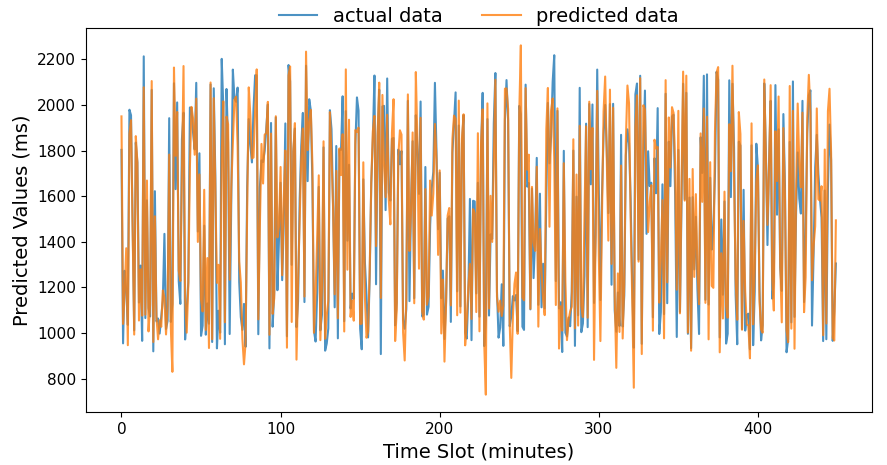}
\caption{Accurate predictions achieved by MSARS.}
\label{fig:gnn_hig}
\end{figure}

Finally, to test MSARS in extremely dynamic microservice environments, we concatenated data from the dataset to generate data with significant fluctuations. MSARS have shown to learn this high volatility, resulting in highly accurate predictions, as shown in the Figure \ref{fig:gnn_hig}.

\subsection{RL Decision with SLO Allocation Strategies}

Incorporating the generated SLO resource allocation strategy as part of the state and reward in RL helps the agent better understand the conditions for resource scaling in the current environment. The local SLO resource values can guide the agent to determine the optimal horizontal scaling thresholds when scaling resources, maximizing resource savings while ensuring QoS meet SLO.

\begin{figure}[htbp]
\centering
\includegraphics[width=0.6\linewidth]{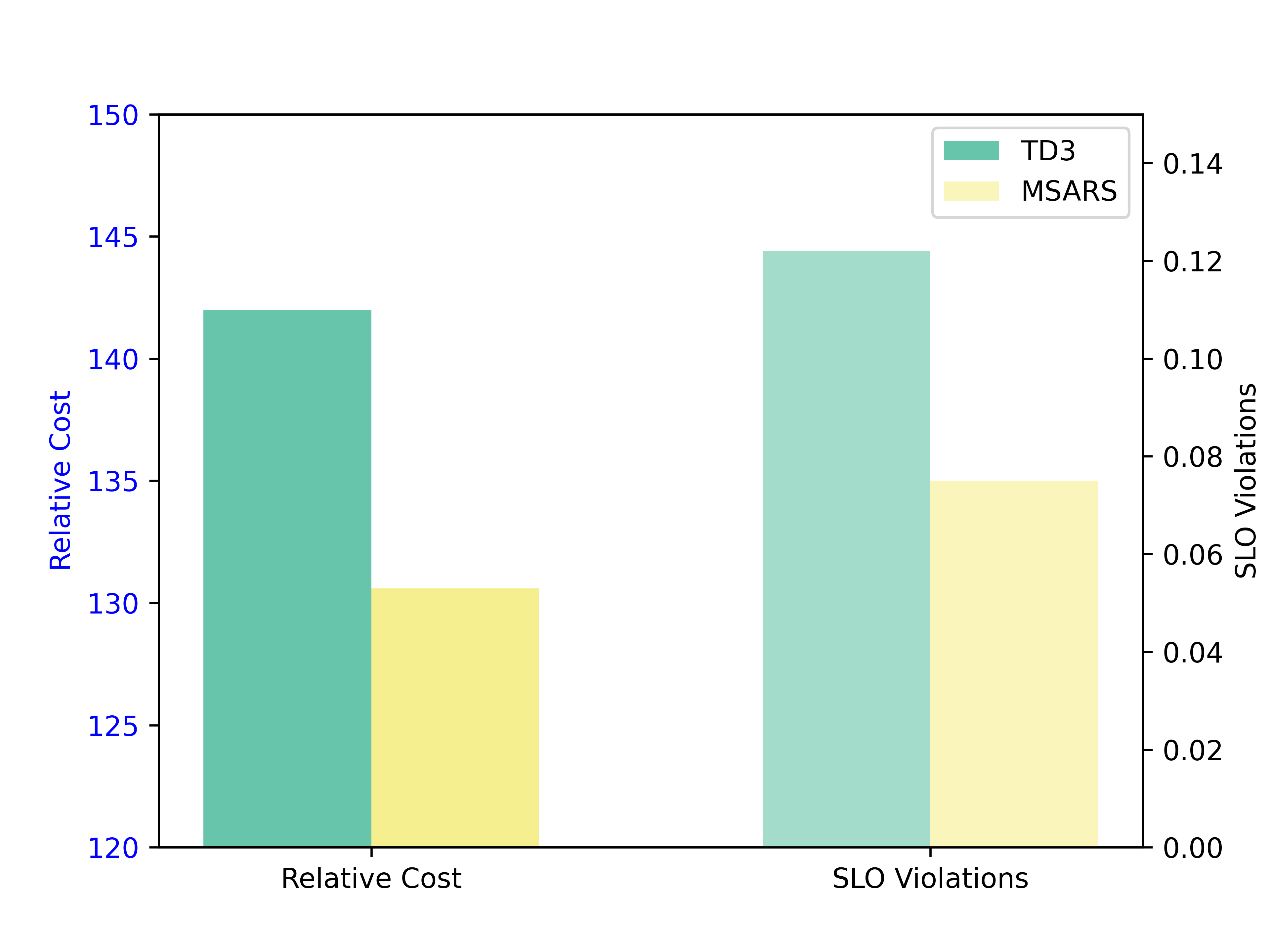}
\caption{Cost and SLO violation comparison with TD3.}
\label{fig:compare}
\end{figure}

To this end, we compared the relative resource cost (normalized with the maximum resource usage) and SLO violations of MSARS with the TD3 algorithm \cite{DRPC}, as shown in the Figure \ref{fig:compare}. We can see that compared to TD3, the inclusion of the SLO resource allocation policy makes the resource scheduling of MSARS always make decisions in a way that satisfies the SLO demand, resulting in an 38\% reduction in the SLO default rate of the whole system compared to the TD3 default rate. Due the agent makes decisions on vertical scaling by adjusting the size of microservice instances, this results in MSARS regenerating the SLO resource allocation strategy, which undoubtedly consumes some resources. If container vertical scaling occurs very frequently, it is undoubtedly a waste of resources. Therefore, MSARS encourages more horizontal scaling of instances and only opts for vertical scaling when fine-grained load changes require maintaining high CPU utilization, so with regard to savings in relative resource cost, the results were not as high as initially expected, with 8\% reduction 8\% in  resource consumption compared to the TD3.

To summarize, our approach can quickly adapt to new environments and generate new SLO resource allocation schemes. The model's convergence time is reduced by 40\% compared to baseline methods, since the \textit{Meta Learner} optimises its initial parameters by training on multiple tasks, it enables effective learning of new tasks from this set of initial parameters with a small number of gradient updating steps and a small amount of data, enabling effective adaptation to new environments and applications. SLO-oriented resource scaling effectively ensures the QoS of microservices, significantly reduces SLO violation rates, and decreases system resource consumption.

\section{Related Work}

% Current research on SLO allocation mainly focus on techniques based on SLO-oriented allocation, reinforcement learning, and meta-learning.
There is currently a large body of research on the auto-scaling of microservice resources that focuses on SLO to ensure QoS. This includes methods that allocate SLO resources as system resources for scaling microservices to maintain QoS, frameworks that use reinforcement learning to automatically scale resources to meet SLO, and approaches using meta-learning to adapt to environmental changes to ensure SLO compliance.

\subsection{SLO allocation ensures SLO}

Optimization-based SLO allocation aims to efficiently distribute resources using mathematical or machine learning techniques to achieve predefined performance targets while minimizing costs and ensuring service quality, which treat SLO allocation as an optimization problem.

Mirhosseini et al. \cite{Parslo} proposed Parslo, an optimization method based on gradient descent, which minimizes deployment costs while ensuring end-to-end SLO fulfillment. However, its response time during transient traffic changes is longer compared to some centralized machine learning methods. For instance, Luo et al. \cite{Erms} proposed Erms, an efficient resource management system that enhances resource utilization and reduces SLO violation probabilities through optimized scheduling strategies and resource allocation. Despite requiring complex dependency graph analysis and multi-tenant priority scheduling, Erms achieves efficient resource management and SLO assurance in multi-tenant environments by leveraging tail latency piecewise linear function modeling and priority scheduling strategies.

\subsection{Reinforcement-Learning ensures SLO}

RL-based SLO allocatiopn \cite{hyscale} enhances system adaptability to uncertainty and change through trial-and-error learning. These methods dynamically adjust resource allocation based on real-time workload and environmental changes, aiming to maximize system performance or meet specific SLOs.

Zhang et al. \cite{A-SARSA} proposed A-SARSA, a predictive container auto-scaling algorithm based on RL, combining SARSA and ARIMA models to dynamically forecast workload, thereby accelerating the convergence speed of RL. Despite potential initial performance challenges during cold starts, experimental results demonstrate that the algorithm significantly reduces SLO violation rates while excelling in timeliness and accuracy of decision-making. Wang et al. \cite{DeepScaling} introduced DeepScaling, employing deep learning and RL methods to achieve more accurate load prediction and CPU utilization estimation, thus improving resource efficiency and reducing costs. However, this method requires substantial computational resources for model training and inference, making it less suitable for small-scale systems. Some works also consider anomaly detection, such as ADRL proposed by Kardani-Moghaddam et al. \cite{ADRL}, which integrates anomaly detection with RL to optimize resource scaling in cloud environments. However, due to the complexity of resource adjustment based on multi-tiered decision-making and environmental feedback, the model may experience higher false positives during state transitions, potentially affecting practical application effectiveness.

\subsection{Meta-Learning ensures SLO}

Meta learning learns how to rapidly adapt to new workload patterns and environmental changes \cite{metaslo}, optimizing auto-scaling strategies to ensure the achievement of SLOs, which shows advantage in real-time resource allocation.

Qiu et al. \cite{AWARE} presented AWARE, an automated workload scaling framework based on meta-learning. Although this framework is not sufficiently general-purpose and requires specific design for particular workloads, as well as time-consuming retraining for new changes, AWARE achieves rapid adaptation to diverse workloads and environments through the integration of meta-learning and RL. Qu et al. \cite{DMRO} designed DMRO, combining deep RL with meta-learning to enhance adaptability and efficiency in dynamic environments, demonstrating strong intelligent decision-making capabilities and efficiency. However, its computational complexity is high, and initial training may require a longer time and more training steps to converge. Xue et al. \cite{AMetaReinforcement} developed a predictive auto-scaling method based on meta reinforcement learning, significantly improving the accuracy and efficiency of cloud resource allocation by integrating deep learning and neural process models. However, it may not perform optimally in systems with insufficient or low-quality data.

Compared to baselines, MSARS focuses on rapidly adapting to the  microservice environment to generate SLO allocation strategies. By analyzing the three features mentioned above as inputs, it can accurately allocate end-to-end SLO resources, saving significant iterative computation time and resources. Unlike other SLO resource allocation methods, MSARS extends them from static to dynamic environments, quickly responding to microservice application updates, chain changes, and the addition of applications. This results in a broader range of application scenarios and more effective and stable allocation strategies. Additionally, embedding the SLO allocation strategy into the decision-making of RL agents makes resource scaling decisions more aligned with the characteristics of microservice resource scaling, optimizing system resource usage.

\section{Conclusion and Future Work}
Our work is dedicated to the rapid allocation of partial SLO resources in dynamic environments, guiding the resource scaling of microservice applications. We designed, MSARS, a comprehensive resource management framework based on meta-learning and reinforcement learning within a microservice architecture. This framework can quickly and accurately obtain near optimal SLO resource allocation schemes, maximizing system resource savings while ensuring QoS to meet SLO constraints. Moreover, we thoroughly analyzed the factors affecting SLO resource allocation in dynamic environments, and examined the changes in microservice instance LLP characteristics caused by instance size variations and version updates, which contributes to the design of the \textit{Meta Learner}. This allows MSARS to quickly obtain optimal SLO resource allocation strategies in dynamic environments and integrate them into RL to guide scaling decisions. Compared with baselines, MSARS reduces the time required to adapt to new environments by 40\%, lowers the SLO violation by 38\%, and decreases resource consumption by 8\%. In future work, we will focus on optimizing the deep integration between SLO resources and reinforcement learning agents, and extend the application of meta-learning. This will enable the scaling framework to effectively handle highly volatile and irregular burst loads, thereby addressing issues related to SLO violations.

\section*{Acknowledgments} This work is supported by National Key R \& D Program of China (No. 2021YFB3300200), and the National Natural Science Foundation of China (No. 62072451, 62102408, 92267105), Guangdong Basic and Applied Basic Research Foundation (No. 2024A1515010251, 2023B1515130002). %Guangdong Special Support Plan (No. 2021TQ06X990), Shenzhen Basic Research Program (No. JCYJ20220818101610023), Shenzhen Industrial Application Projects of undertaking the National key R \& D Program of China (No. CJGJZD20210408091600002).

\bibliographystyle{unsrt}
\bibliography{sample-based}

\end{document}